\documentclass[preprint,showpacs,preprintnumbers,amsmath,amssymb]{revtex4}
\usepackage{graphicx}
\usepackage{feynmf}
\newcommand{\bfalpha}{{\mbox{\boldmath$\alpha$}}}
\newcommand{\bfbeta}{{\mbox{\boldmath$\beta$}}}
\newcommand{\bfgamma}{{\mbox{\boldmath$\gamma$}}}

\newcommand{\kk}{{\bf k}}

\newcommand{\rr}{{\bf r}}

\newcommand{\dd}{{\mbox{d}}}

\newcommand{\be}{\begin{equation}}
\newcommand{\ee}{\end{equation}}
\newcommand{\ba}{\begin{eqnarray}}
\newcommand{\ea}{\end{eqnarray}}
\newcommand{\bse}{\begin{subequations}}
\newcommand{\ese}{\end{subequations}}
\newcommand{\beq}{\begin{eqnarray}}
\newcommand{\eeq}{\end{eqnarray}}

\begin{document}
\begin{fmffile}{ERM_mf}

\title{Euclidean random matrix theory: low-frequency non-analyticities
and Rayleigh scattering}
\author{Carl Ganter}
\affiliation{Institut f\"ur R\"ontgendiagnostik, Technische
Universit\"at M\"unchen, Ismaninger Str. 22, D-81675 M\"unchen}
\author{Walter Schirmacher}
\affiliation{Institut f\"ur Physik,
Universit\"at Mainz, D-55099 Mainz, Germany}
\affiliation{Physik-Department E13, Technische Universit\"at
M\"unchen, D-85747, Garching, Germany}

\begin{abstract}
By calculating all terms of the high-density expansion of
the euclidean random matrix theory 
(up to second-order in the inverse density)
for the vibrational spectrum
of a topologically disordered system we show that
the low-frequency behavior of the self energy is given
by $\Sigma(k,z)\propto k^2z^{d/2}$ and not
 $\Sigma(k,z)\propto k^2z^{(d-2)/2}$, as claimed previously.
This implies the presence of Rayleigh scattering and
long-time tails of the velocity autocorrelation function
of the analogous diffusion problem of the
form $Z(t)\propto t^{-(d+2)/2}$.

\end{abstract}
\maketitle

\section{Introduction}
Rayleigh scattering \cite{rayleigh}, i.e. the fact
that the mean-free path of weakly scattered waves
varies as $\omega^{-(d+1)}$ in a $d$-dimensional
disordered medium as $\omega\rightarrow 0$, is widely
believed to be a general property of quenched disordered
matter. However, recently it has been claimed
\cite{parisi}
that
a harmonic system with displacements $u_i(t)$ obeying
\be\label{eqmo1}
\frac{\dd^2}{\dd t^2}u_i(t)=-\sum_jt_{ij}\big(u_i(t)-u_j(t)\big)\, ,
\ee
where $i$ and $j$ denote random sites $\rr_{i,j}$ in
$d$-dimensional space, 
would have wave-like excitations, which have a line-width
(inverse mean-free path), varying with 
$\omega^2$ instead of $\omega^4$ in $d=3$. 
$t_{ij}$ are force constants, divided the mass at the
node $i$, which are assumed to depend on the distance,
i.e. $t_{ij}=t(r_{ij})$.
The claim of absence of Rayleigh scattering
had been substantiated
by a high-density expansion and a diagrammatical analysis
\cite{parisi}. This claim
is not
only astonishing with respect to the mentioned general
view on waves in disordered media, but it is also in
contradiction with 
the known analytic properties of the analogous
diffusion system. If one replaces the double time derivative
in (\ref{eqmo1}) by a single one, 
one obtains the equation
of a random walk among the sites $i,j$: 
\be\label{eqmo2}
\frac{\dd}{\dd t}n_i(t)=-\sum_jt_{ij}\big(n_i(t)-n_j(t)\big)\, ,
\ee
where $n_i(t)$ give the odds for the walker to be at
$i$ at time $t$ and $t_{ij}=t(r_{ij})$ is the hopping
probability per unit time.
Eq. (\ref{eqmo2}) describes e.g. the 
motion of electrons hopping among
shallow impurities
in a semiconductor \cite{ma60,ahl71,se79,bot85}. 
Such a random walk
is known \cite{haus83} to exhibit a long-time tail
of the velocity-autocorrelation function (VAF) varying
as $Z(t)\propto t^{-(d+2)/2}$ for $t\rightarrow \infty$
\cite{haus83, ernst84}, a feature shared with
Lorentz models \cite{ernst71,ernst84,hof06}. In fact,
the Laplace transform of the VAF is the frequency-dependent
diffusivity $D(z=i\omega+\epsilon)$, which has, according to
the Tauberian theorems \cite{feller50} a low-frequency
singularity $D(z)\rightarrow z^{d/2}, |z|\rightarrow 0$.
Now, in the analogous vibrational problem this quantity
corresponds to the square of a frequency-dependent
sound velocity $D(z=-\omega^2+i\epsilon)=v^2(z)$. The imaginary
part $v''(\omega)$
of the latter is related to the mean-free part via
\be
\frac{1}{\ell(\omega)}=\frac{2\omega v''(\omega)}{|v(z)|^2}
\ee
This gives $\ell\propto \omega^{-(d+1)}$, i.e. Rayleigh
scattering. We conclude that the long-time tail
of the VAF in the diffusion problem is mathematically
equivalent to the Rayleigh-scattering property. 

In the following we set the dimensionality
$d=3$ and calculate explicitly all 
irreducible diagrams (self-energy diagrams)
up to second order in the inverse
density $\rho^{-1}=V/N$, where $N$ is the number of
sites and $V$ the volume. We show that
to this order the self energy is proportional
to $k^2z^{3/2}$, $z=i\omega+\epsilon$ (diffusion)
or $z=-\omega^2+i\epsilon$ (sound). and not
as claimed in Refs. \cite{parisi}
$\propto k^2z^{1/2}$. We also show, why the
so-called cactus approximation for a self-consistent
theory erroneously leads to a non-analyticity
$z^{1/2}$ instead of $z^{3/2}$.

\section{Formalism}
 
As in refs. \cite{parisi} we start from
a high-frequency
($z$) and high-density ($\rho=N/V$) expansion of the averaged propagator

\beq\label{prop}
&&G(\kk,z)=\frac{1}{N}
\sum_{mn}\bigg\langle
e^{i\kk\rr_{mn}}[z\mathbf{1}-\mathbf{M}]^{-1}_{mn}\bigg\rangle
=\frac{1}{z}
\nonumber\\
&&+\sum_{p=1}^\infty\frac{1}{z^{p+1}}
\frac{1}{N}\sum_{i_0\dots i_p}
\bigg\langle
e^{i\kk\rr_{i_0i_1}}
M_{i_0i_1}
\dots
e^{i\kk\rr_{i_{p-1}i_p}}
M_{i_{p-1}i_p}
\bigg\rangle\nonumber
\eeq

Here {\bf M}
is a matrix with off-diagonal elements
\mbox{$M_{ij}=t_{ij}$} and diagonal elements \mbox{$M_{ii}=-\sum_{\ell\neq i}
t_{i\ell}$.} 
$t(k)=t(\kk)$ is the $d$-dimensional Fourier
transform of $t(r)$. 

The configurationally averaged Green's function can now
be expressed in terms of
the irreducible self energy~$\Sigma\left(k,z\right)$ as follows

\begin{equation}
G\left(k,z\right) = 
\frac{1}{z - \rho\left[\,t\left(k\right) - t\left(0\right)\,\right] - 
\Sigma\left(k,z\right)}
\stackrel{k\rightarrow 0}{=}
\frac{1}{z+D(z)k^2}
\end{equation}

The frequency-dependent diffusivity/sound velocity is given by

\be
D(z)=v(z)^2=-\frac{1}{2}\frac{\partial^2}{\partial k^2}
\bigg[t(k)+\Sigma(k,z)\bigg]_{k\rightarrow 0}
\ee

For simplicity, we assume complete site disorder (i.e. 
the radial pair correlation function \mbox{$g\left(r\right)\equiv1$).}
Therefore $t\left(k\right)$ is simply the Fourier transform of the
transition rate $t\left(r\right)$ \cite{comment1}.

We denote the unrenormalized part of the Green's function by $G_0$:
\begin{equation}
G_0\left(k,z\right) := 
\frac{1}{z - \rho\left[\,t\left(k\right) - t\left(0\right)\,\right]}
\end{equation}

In analogy to the approach in Ref.~\cite{parisi}, it is helpful to consider a
high-density expansion of the propagator, which is in turn determined by an
analogous expansion the self energy~$\Sigma\left(k,z\right)$

\begin{equation}
\Sigma\left(k,z\right) =:
\sum_{n=1}^\infty\,\rho^{-n}\,\Sigma^{(n)}\left(k,z\right),
\end{equation}

As outlined in \cite{parisi}, the index~$n$ counts repetitions of sites in
the high frequency / high density expansion.

In the following, we will derive exact results for $n=1$ and $n=2$.

To this end, we will use diagrammatic representations, to distinguish
topologically different contributions to the self energy.

\begin{itemize}
\item Off-diagonal matrix elements, associated with a site change. Open
  circles will always indicate start and end point of a diagram.
\begin{equation}
t^o_{12} \equiv t\left(r_{12}\right):=
\qquad\qquad
1\quad\parbox{3cm}{
\begin{fmfgraph*}(3,1.5) \fmfpen{thick}
\fmfleft{v1}
\fmfright{v2}
\fmf{fermion}{v1,v2} 
\fmfblob{4thick}{v1}
\fmfblob{4thick}{v2}
\end{fmfgraph*}
}\quad 2,
\end{equation}
\item Part of the diagonal element, associated with
  site~$2$ - its value is determined by the sum rule. Since there is no site
  change, only one open circle is drawn.
\begin{equation}
t^d_{12} \equiv -\,t\left(r_{12}\right) := 
\qquad\qquad
1\quad\parbox{3cm}{
\begin{fmfgraph*}(3,1.5) \fmfpen{thick}
\fmfleft{v1}
\fmfright{v2}
\fmf{scalar}{v1,v2} 
\fmfblob{4thick}{v1}
\end{fmfgraph*}
}\quad 2
\end{equation}
\item The unrenormalized propagator
\begin{equation}
\left[\,G_0\,\right]_{12} \equiv G_0\left(r_{12}\right) := 
\qquad\qquad
1\quad\parbox{3cm}{
\begin{fmfgraph*}(3,1.5) \fmfpen{thick}
\fmfleft{v1}
\fmfright{v2}
\fmf{heavy}{v1,v2} 
\fmfblob{4thick}{v1}
\fmfblob{4thick}{v2}
\end{fmfgraph*}
}\quad 2
\end{equation}
\end{itemize}

Note that the propagator contains a diagonal part (sites~$1$ and $2$
are the same, i.e. the diagram has length zero)

\begin{equation}
{\cal G}_0\left(z\right) := \frac{1}{z + \rho\,t\left(0\right)}
\end{equation}

formally obtained as ${\cal G}_0\left(z\right) =
\lim_{k\to\infty}\,G_0\left(k,z\right)$. In most cases, this requires no
special attention. Exceptions, when these terms need to be explicitly excluded
to preserve irreducibility will be mentioned below.

\section{First-order diagrams:  $\Sigma^{(1)}\left(k,z\right)$}

This case is comparably trivial and requires the addition of four diagrams,
since the first and last connection can refer to an off--diagonal ({\bf O})
or a diagonal ({\bf D}) transition rate.

\subsubsection*{OO:}

\begin{equation}\label{OO_ex}
\parbox{3cm}{
\begin{fmfgraph*}(3,2.5) \fmfpen{thick}
\fmftop{t1} \fmfbottom{b1,b2} 
\fmf{heavy}{b1,b2} 
\fmf{fermion}{t1,b1} 
\fmf{fermion}{b2,t1} 
\fmfblob{4thick}{t1}
\fmfdot{b1,b2}
\fmflabel{{\bf 1}}{t1}
\end{fmfgraph*}}
\qquad
=
\qquad
\rho\int \frac{d{\bf p}}{\left(2\pi\right)^3}\,\,t^2\left(p\right)\,G_0\left(p,z\right) 
\end{equation}

\subsubsection*{OD:}

\begin{equation}\label{OD_ex}
\parbox{3cm}{
\begin{fmfgraph*}(3,2.5) \fmfpen{thick}
\fmftop{t1} \fmfbottom{b1,b2} 
\fmf{heavy}{b1,b2} 
\fmf{fermion}{t1,b1} 
\fmf{scalar}{b2,t1} 
\fmfdot{b1}
\fmfblob{4thick}{t1,b2}
\fmflabel{{\bf 1}}{t1}
\end{fmfgraph*}}
\qquad
=
\qquad
-\,\rho\int \frac{d{\bf p}}{\left(2\pi\right)^3}\,\, t\left({\bf k}-{\bf
    p}\right)\,t\left(p\right)\,G_0\left(p,z\right) 
\end{equation}

\subsubsection*{DO:}

\begin{equation}\label{DO_ex}
\parbox{3cm}{
\begin{fmfgraph*}(3,2.5) \fmfpen{thick}
\fmftop{t1} \fmfbottom{b1,b2} 
\fmf{heavy}{b1,b2} 
\fmf{scalar}{b1,t1} 
\fmf{fermion}{b2,t1} 
\fmfblob{4thick}{b1,t1}
\fmfdot{b2}
\fmflabel{{\bf 1}}{t1}
\end{fmfgraph*}}
\qquad
=
\qquad
-\,\rho \int \frac{d{\bf p}}{\left(2\pi\right)^3}\,\, t\left({\bf k}-{\bf
    p}\right)\,t\left(p\right)\,G_0\left(p,z\right) 
\end{equation}

\subsubsection*{DD:}

\begin{equation}\label{DD_ex}
\parbox{3cm}{
\begin{fmfgraph*}(3,2.5) \fmfpen{thick}
\fmftop{t1} \fmfbottom{b1,b2} 
\fmf{heavy}{b1,b2} 
\fmf{scalar}{b1,t1} 
\fmf{scalar}{b2,t1} 
\fmfblob{4thick}{b1,b2}
\fmflabel{{\bf 1}}{t1}
\end{fmfgraph*}}
\qquad
=
\qquad
\rho\int \frac{d{\bf p}}{\left(2\pi\right)^3}\,\, t^2\left({\bf k} - {\bf
    p}\right)\,G_0\left(p,z\right) 
\end{equation}

\subsubsection*{Added together:}

\begin{equation}\label{sig_1_ex}
\rho\int \frac{d{\bf p}}{\left(2\pi\right)^3}
\left[\,t\left({\bf k} - {\bf p}\right) - t\left(p\right)\,\right]^2
G_0\left(p,z\right)
\end{equation}

Since we are mainly interested in the imaginary part of

\begin{equation}
\lim_{z\to 0}\,
\lim_{k\to 0}\, \Sigma\left(k,z\right)
\end{equation}

we will have to examine the bracket in Eq.~(\ref{sig_1_ex}) in the limit~$k\to
0$ 

\begin{equation}\label{tkpk}
\lim_{k\to 0}\,
t\left({\bf k} - {\bf p}\right) - t\left(p\right)
= 
\frac{t^\prime\left(p\right)}{p}\cdot{\bf kp} 
+ \frac{1}{2}\,\frac{t^{\prime\prime}\left(p\right)\cdot p -
  t^\prime\left(p\right)}{p^3}
\cdot\left[\,{\bf kp}\,\right]^2
+ \frac{1}{2}\,\frac{t^\prime\left(p\right)}{p}
\cdot k^2
+ {\cal O}\left(k^3\right)
\end{equation}

We therefore obtain

\begin{equation}\label{sig_1_ex_appr}
\lim_{p\to 0}\,\lim_{k\to 0}\,
\left[\,t\left({\bf k} - {\bf p}\right) - t\left(p\right)\,\right]^2
= 
c\cdot\underbrace{\left[\,{\bf kp}\,\right]^2}_{\propto \,p^2}
+ {\cal O}\left(k^3,\,p^3\right)
\end{equation}

with some usually nonzero constant~$c$. With the additional $p^2$ factor from the
threedimensional integral, we obtain from the diffusion pole of
$G_0\left(k,z\right)$ 

\begin{equation}
\lim_{z\to
    0}\,\lim_{k\to 0}\,{\rm
    Im}\left[\,\Sigma^{(1)}\left(k,z\right)\,\right]\propto z^{3/2}k^2
\end{equation}

\section{Second-order diagrams: $\Sigma^{(2)}\left(k,z\right)$}

This case is considerably more complex. It turns out to be advantegeous to
consider topologically different groups of irreducible diagrams separately:

\begin{description}
\item[$\Sigma_\alpha^{(2)}\left(k,z\right)$: ] $(1 * 2 \cdots 2 \cdots 1)$
\item[$\Sigma_\beta^{(2)}\left(k,z\right)$: ]  $(1\,2 \cdots 2 \cdots 1)$
\item[$\Sigma_\gamma^{(2)}\left(k,z\right)$: ]  $(1 * 2 \cdots 1 \cdots 2)$
\item[$\Sigma_\delta^{(2)}\left(k,z\right)$: ]  $(1\, 2 \cdots 1 \cdots 2)$
\item[$\Sigma_\varepsilon^{(2)}\left(k,z\right)$: ]  $(1 \cdots 1 \cdots 1)$
\end{description}

Unlike in $\cdots$ at least one additional site index (differnt from $1$ and $2$)
needs to be contained in $*$.

The complete and exact second-order self energy is then just the sum of
these partial contributions:

\begin{equation}\label{sig_2_split}
\Sigma^{(2)}\left(k,z\right) = 
\Sigma_\alpha^{(2)}\left(k,z\right) +
\Sigma_\beta^{(2)}\left(k,z\right) +
\Sigma_\gamma^{(2)}\left(k,z\right) +
\Sigma_\delta^{(2)}\left(k,z\right) +
\Sigma_\varepsilon^{(2)}\left(k,z\right)
\end{equation}

\subsection{$\Sigma_\alpha^{(2)}\left(k,z\right)$: Irreducible Diagrams $(1 * 2 \cdots 2 \cdots 1)$}

Here, 16~diagrams need to be distinguished \cite{comment2}:

\subsubsection*{OOOO:}

\begin{equation}\label{OOOO_ex}
\parbox{4.5cm}{
\begin{fmfgraph*}(4.5,3) \fmfpen{thick}
\fmfsurround{z3,d1,z4,s1,z1,z2,d2}
\fmf{fermion}{s1,z1} 
\fmf{fermion}{z4,s1} 
\fmf{heavy}{z1,z2} 
\fmf{heavy,right}{s2,z3} 
\fmf{fermion,right}{z3,s2} 
\fmf{fermion}{z2,s2} 
\fmf{heavy}{s2,z4} 
\fmfdot{z1,z2,z3,z4,s2}
\fmfblob{4thick}{s1}
\fmflabel{{\bf 1}}{s1}
\fmflabel{{\bf 2}}{s2}
\end{fmfgraph*}}
\quad
=
\quad
\rho^2\int 
d{\bf p}\,d{\bf q}\,
\,t^3\left(p\right)
\,t\left(q\right)
\,G^2_0\left(p\right) 
\,G_0\left(q\right) 
\end{equation}

\subsubsection*{OOOD:}

\begin{equation}\label{OOOD_ex}
\parbox{4.5cm}{
\begin{fmfgraph*}(4.5,3) \fmfpen{thick}
\fmfsurround{z3,d1,z4,s1,z1,z2,d2}
\fmf{fermion}{s1,z1} 
\fmf{scalar}{z4,s1} 
\fmf{heavy}{z1,z2} 
\fmf{heavy,right}{s2,z3} 
\fmf{fermion,right}{z3,s2} 
\fmf{fermion}{z2,s2} 
\fmf{heavy}{s2,z4} 
\fmfdot{z1,z2,z3,s2}
\fmfblob{4thick}{s1,z4}
\fmflabel{{\bf 1}}{s1}
\fmflabel{{\bf 2}}{s2}
\end{fmfgraph*}}
\quad
=
\quad
-\,\rho^2\int 
d{\bf p}\,d{\bf q}\,
\,t\left({\bf k} - {\bf p}\right)
\,t^2\left(p\right)
\,t\left(q\right)
\,G^2_0\left(p\right) 
\,G_0\left(q\right) 
\end{equation}

\subsubsection*{OODO:}

\begin{equation}\label{OODO_ex}
\parbox{4.5cm}{
\begin{fmfgraph*}(4,3) \fmfpen{thick}
\fmfsurround{d1,z3,z4,s1,z1,z2,s2}
\fmf{fermion}{s1,z1} 
\fmf{heavy}{z1,z2} 
\fmf{fermion}{z2,s2} 
\fmf{heavy}{s2,z3} 
\fmf{scalar,left}{z3,s2}
\fmf{heavy}{z3,z4} 
\fmf{fermion}{z4,s1} 
\fmfdot{z1,z2,z3,z4,s2}
\fmfblob{4thick}{s1}
\fmflabel{{\bf 1}}{s1}
\fmflabel{{\bf 2}}{s2}
\end{fmfgraph*}}
\quad
=
\quad
-\,\rho^2\int 
d{\bf p}\,d{\bf q}\,
\,t^3\left(p\right)
\,t\left({\bf p} - {\bf q}\right)
\,G^2_0\left(p\right) 
\,G_0\left(q\right) 
\end{equation}

\subsubsection*{OODD:}

\begin{equation}\label{OODD_ex}
\parbox{4.5cm}{
\begin{fmfgraph*}(4,3) \fmfpen{thick}
\fmfsurround{d1,z3,z4,s1,z1,z2,s2}
\fmf{fermion}{s1,z1} 
\fmf{heavy}{z1,z2} 
\fmf{fermion}{z2,s2} 
\fmf{heavy}{s2,z3} 
\fmf{scalar,left}{z3,s2}
\fmf{heavy}{z3,z4} 
\fmf{scalar}{z4,s1} 
\fmfdot{z1,z2,z3,s2}
\fmfblob{4thick}{s1,z4}
\fmflabel{{\bf 1}}{s1}
\fmflabel{{\bf 2}}{s2}
\end{fmfgraph*}}
\quad
=
\quad
\rho^2\int 
d{\bf p}\,d{\bf q}\,
\,t\left({\bf k} - {\bf p}\right)
\,t^2\left(p\right)
\,t\left({\bf p} - {\bf q}\right)
\,G^2_0\left(p\right) 
\,G_0\left(q\right) 
\end{equation}

\subsubsection*{ODOO:}

\begin{equation}\label{ODOO_ex}
\parbox{4.5cm}{
\begin{fmfgraph*}(4.5,3) \fmfpen{thick}
\fmfsurround{z3,s2,z4,s1,z1,z2}
\fmf{fermion}{s1,z1} 
\fmf{fermion}{z4,s1} 
\fmf{fermion}{z3,s2} 
\fmf{heavy}{z1,z2} 
\fmf{scalar}{z2,s2} 
\fmf{heavy}{z2,z3} 
\fmf{heavy}{s2,z4} 
\fmfdot{z1,z2,z3,z4,s2}
\fmfblob{4thick}{s1}
\fmflabel{{\bf 1}}{s1}
\fmflabel{{\bf 2}}{s2}
\end{fmfgraph*}}
\quad
=
\quad
-\,\rho^2\int 
d{\bf p}\,d{\bf q}\,
\,t^2\left(p\right)
\,t\left({\bf p} - {\bf q}\right)
\,t\left(q\right)
\,G^2_0\left(p\right) 
\,G_0\left(q\right) 
\end{equation}

\subsubsection*{ODOD:}

\begin{equation}\label{ODOD_ex}
\parbox{4.5cm}{
\begin{fmfgraph*}(4.5,3) \fmfpen{thick}
\fmfsurround{z3,s2,z4,s1,z1,z2}
\fmf{fermion}{s1,z1} 
\fmf{scalar}{z4,s1} 
\fmf{fermion}{z3,s2} 
\fmf{heavy}{z1,z2} 
\fmf{scalar}{z2,s2} 
\fmf{heavy}{z2,z3} 
\fmf{heavy}{s2,z4} 
\fmfdot{z1,z2,z3,s2}
\fmfblob{4thick}{s1,z4}
\fmflabel{{\bf 1}}{s1}
\fmflabel{{\bf 2}}{s2}
\end{fmfgraph*}}
\quad
=
\quad
\rho^2\int 
d{\bf p}\,d{\bf q}\,
\,t\left({\bf k} - {\bf p}\right)
\,t\left(p\right)
\,t\left({\bf p} - {\bf q}\right)
\,t\left(q\right)
\,G^2_0\left(p\right) 
\,G_0\left(q\right) 
\end{equation}

\subsubsection*{ODDO:}

\begin{equation}\label{ODDO_ex}
\parbox{5cm}{
\begin{fmfgraph*}(4.5,3) \fmfpen{thick}
\fmfsurround{s2,z3,z4,s1,z1,z2}
\fmf{fermion}{s1,z1} 
\fmf{fermion}{z4,s1} 
\fmf{scalar}{z2,s2} 
\fmf{scalar}{z3,s2} 
\fmf{heavy}{z1,z2} 
\fmf{heavy}{z2,z3} 
\fmf{heavy}{z3,z4} 
\fmfdot{z1,z2,z3,z4}
\fmfblob{4thick}{s1}
\fmflabel{{\bf 1}}{s1}
\fmflabel{{\bf 2}}{s2}
\end{fmfgraph*}}
\quad
=
\quad
\rho^2\int 
d{\bf p}\,d{\bf q}\,
\,t^2\left(p\right)
\,t^2\left({\bf p} - {\bf q}\right)
\,G^2_0\left(p\right) 
\,G_0\left(q\right) 
\end{equation}

\subsubsection*{ODDD:}

\begin{equation}\label{ODDD_ex}
\parbox{5cm}{
\begin{fmfgraph*}(4.5,3) \fmfpen{thick}
\fmfsurround{s2,z3,z4,s1,z1,z2}
\fmf{fermion}{s1,z1} 
\fmf{scalar}{z4,s1} 
\fmf{scalar}{z2,s2} 
\fmf{scalar}{z3,s2} 
\fmf{heavy}{z1,z2} 
\fmf{heavy}{z2,z3} 
\fmf{heavy}{z3,z4} 
\fmfdot{z1,z2,z3}
\fmfblob{4thick}{s1,z4}
\fmflabel{{\bf 1}}{s1}
\fmflabel{{\bf 2}}{s2}
\end{fmfgraph*}}
\quad
=
\quad
-\,\rho^2\int 
d{\bf p}\,d{\bf q}\,
\,t\left({\bf k} - {\bf p}\right)
\,t\left(p\right)
\,t^2\left({\bf p} - {\bf q}\right)
\,G^2_0\left(p\right) 
\,G_0\left(q\right) 
\end{equation}

\subsubsection*{DOOO:}

\begin{equation}\label{DOOO_ex}
\parbox{4.5cm}{
\begin{fmfgraph*}(4.5,3) \fmfpen{thick}
\fmfsurround{z3,d1,z4,s1,z1,z2,d2}
\fmf{scalar}{z1,s1} 
\fmf{fermion}{z4,s1} 
\fmf{heavy}{z1,z2} 
\fmf{heavy,right}{s2,z3} 
\fmf{fermion,right}{z3,s2} 
\fmf{fermion}{z2,s2} 
\fmf{heavy}{s2,z4} 
\fmfdot{z1,z2,z3,z4,s2}
\fmfblob{4thick}{s1,z1}
\fmflabel{{\bf 1}}{s1}
\fmflabel{{\bf 2}}{s2}
\end{fmfgraph*}}
\quad
=
\quad
-\,\rho^2\int 
d{\bf p}\,d{\bf q}\,
\,t\left({\bf k} - {\bf p}\right)
\,t^2\left(p\right)
\,t\left(q\right)
\,G^2_0\left(p\right) 
\,G_0\left(q\right) 
\end{equation}

\subsubsection*{DOOD:}

\begin{equation}\label{DOOD_ex}
\parbox{4.5cm}{
\begin{fmfgraph*}(4.5,3) \fmfpen{thick}
\fmfsurround{z3,d1,z4,s1,z1,z2,d2}
\fmf{scalar}{z1,s1} 
\fmf{scalar}{z4,s1} 
\fmf{heavy}{z1,z2} 
\fmf{heavy,right}{s2,z3} 
\fmf{fermion,right}{z3,s2} 
\fmf{fermion}{z2,s2} 
\fmf{heavy}{s2,z4} 
\fmfdot{z1,z2,z3,s2}
\fmfblob{4thick}{z1,z4}
\fmflabel{{\bf 1}}{s1}
\fmflabel{{\bf 2}}{s2}
\end{fmfgraph*}}
\quad
=
\quad
\rho^2\int 
d{\bf p}\,d{\bf q}\,
\,t^2\left({\bf k} - {\bf p}\right)
\,t\left(p\right)
\,t\left(q\right)
\,G^2_0\left(p\right) 
\,G_0\left(q\right) 
\end{equation}

\subsubsection*{DODO:}

\begin{equation}\label{DODO_ex}
\parbox{4.5cm}{
\begin{fmfgraph*}(4,3) \fmfpen{thick}
\fmfsurround{d1,z3,z4,s1,z1,z2,s2}
\fmf{scalar}{z1,s1} 
\fmf{heavy}{z1,z2} 
\fmf{fermion}{z2,s2} 
\fmf{heavy}{s2,z3} 
\fmf{scalar,left}{z3,s2}
\fmf{heavy}{z3,z4} 
\fmf{fermion}{z4,s1} 
\fmfdot{z1,z2,z3,z4,s2}
\fmfblob{4thick}{s1,z1}
\fmflabel{{\bf 1}}{s1}
\fmflabel{{\bf 2}}{s2}
\end{fmfgraph*}}
\quad
=
\quad
\rho^2\int 
d{\bf p}\,d{\bf q}\,
\,t\left({\bf k} - {\bf p}\right)
\,t^2\left(p\right)
\,t\left({\bf p} - {\bf q}\right)
\,G^2_0\left(p\right) 
\,G_0\left(q\right) 
\end{equation}

\subsubsection*{DODD:}

\begin{equation}\label{DODD_ex}
\parbox{4.5cm}{
\begin{fmfgraph*}(4,3) \fmfpen{thick}
\fmfsurround{d1,z3,z4,s1,z1,z2,s2}
\fmf{scalar}{z1,s1} 
\fmf{heavy}{z1,z2} 
\fmf{fermion}{z2,s2} 
\fmf{heavy}{s2,z3} 
\fmf{scalar,left}{z3,s2}
\fmf{heavy}{z3,z4} 
\fmf{scalar}{z4,s1} 
\fmfdot{z1,z2,z3,s2}
\fmfblob{4thick}{z1,z4}
\fmflabel{{\bf 1}}{s1}
\fmflabel{{\bf 2}}{s2}
\end{fmfgraph*}}
\quad
=
\quad
-\,\rho^2\int 
d{\bf p}\,d{\bf q}\,
\,t^2\left({\bf k} - {\bf p}\right)
\,t\left(p\right)
\,t\left({\bf p} - {\bf q}\right)
\,G^2_0\left(p\right) 
\,G_0\left(q\right) 
\end{equation}

\subsubsection*{DDOO:}

\begin{equation}\label{DDOO_ex}
\parbox{4.5cm}{
\begin{fmfgraph*}(4.5,3) \fmfpen{thick}
\fmfsurround{z3,s2,z4,s1,z1,z2}
\fmf{scalar}{z1,s1} 
\fmf{fermion}{z4,s1} 
\fmf{fermion}{z3,s2} 
\fmf{heavy}{z1,z2} 
\fmf{scalar}{z2,s2} 
\fmf{heavy}{z2,z3} 
\fmf{heavy}{s2,z4} 
\fmfdot{z1,z2,z3,z4,s2}
\fmfblob{4thick}{s1,z1}
\fmflabel{{\bf 1}}{s1}
\fmflabel{{\bf 2}}{s2}
\end{fmfgraph*}}
\quad
=
\quad
\rho^2\int 
d{\bf p}\,d{\bf q}\,
\,t\left({\bf k} - {\bf p}\right)
\,t\left(p\right)
\,t\left({\bf p} - {\bf q}\right)
\,t\left(q\right)
\,G^2_0\left(p\right) 
\,G_0\left(q\right) 
\end{equation}

\subsubsection*{DDOD:}

\begin{equation}\label{DDOD_ex}
\parbox{4.5cm}{
\begin{fmfgraph*}(4.5,3) \fmfpen{thick}
\fmfsurround{z3,s2,z4,s1,z1,z2}
\fmf{scalar}{z1,s1} 
\fmf{scalar}{z4,s1} 
\fmf{fermion}{z3,s2} 
\fmf{heavy}{z1,z2} 
\fmf{scalar}{z2,s2} 
\fmf{heavy}{z2,z3} 
\fmf{heavy}{s2,z4} 
\fmfdot{z1,z2,z3,s2}
\fmfblob{4thick}{z1,z4}
\fmflabel{{\bf 1}}{s1}
\fmflabel{{\bf 2}}{s2}
\end{fmfgraph*}}
\quad
=
\quad
-\,\rho^2\int 
d{\bf p}\,d{\bf q}\,
\,t^2\left({\bf k} - {\bf p}\right)
\,t\left({\bf p} - {\bf q}\right)
\,t\left(q\right)
\,G^2_0\left(p\right) 
\,G_0\left(q\right) 
\end{equation}

\subsubsection*{DDDO:}

\begin{equation}\label{DDDO_ex}
\parbox{5cm}{
\begin{fmfgraph*}(4.5,3) \fmfpen{thick}
\fmfsurround{s2,z3,z4,s1,z1,z2}
\fmf{scalar}{z1,s1} 
\fmf{fermion}{z4,s1} 
\fmf{scalar}{z2,s2} 
\fmf{scalar}{z3,s2} 
\fmf{heavy}{z1,z2} 
\fmf{heavy}{z2,z3} 
\fmf{heavy}{z3,z4} 
\fmfdot{z1,z2,z3,z4}
\fmfblob{4thick}{s1,z1}
\fmflabel{{\bf 1}}{s1}
\fmflabel{{\bf 2}}{s2}
\end{fmfgraph*}}
\quad
=
\quad
-\,\rho^2\int 
d{\bf p}\,d{\bf q}\,
\,t\left({\bf k} - {\bf p}\right)
\,t\left(p\right)
\,t^2\left({\bf p} - {\bf q}\right)
\,G^2_0\left(p\right) 
\,G_0\left(q\right) 
\end{equation}

\subsubsection*{DDDD:}

\begin{equation}\label{DDDD_ex}
\parbox{5cm}{
\begin{fmfgraph*}(4.5,3) \fmfpen{thick}
\fmfsurround{s2,z3,z4,s1,z1,z2}
\fmf{scalar}{z1,s1} 
\fmf{scalar}{z4,s1} 
\fmf{scalar}{z2,s2} 
\fmf{scalar}{z3,s2} 
\fmf{heavy}{z1,z2} 
\fmf{heavy}{z2,z3} 
\fmf{heavy}{z3,z4} 
\fmfdot{z1,z2,z3}
\fmfblob{4thick}{z1,z4}
\fmflabel{{\bf 1}}{s1}
\fmflabel{{\bf 2}}{s2}
\end{fmfgraph*}}
\quad
=
\quad
\rho^2\int 
d{\bf p}\,d{\bf q}\,
\,t^2\left({\bf k} - {\bf p}\right)
\,t^2\left({\bf p} - {\bf q}\right)
\,G^2_0\left(p\right) 
\,G_0\left(q\right) 
\end{equation}

\subsubsection*{Added together:}

\begin{eqnarray}
\lefteqn{\Sigma_\alpha^{(2)}\left(k,z\right) =}
\\\nonumber
&&
\rho^2\int
d{\bf p}\,d{\bf q}\,
\left[\,t\left({\bf k} - {\bf p}\right) - t\left(p\right)\,\right]^2
\,\left[\,t\left({\bf p} - {\bf q}\right) - t\left(p\right)\,\right]
\,\left[\,t\left({\bf p} - {\bf q}\right) - t\left(q\right)\,\right]
\, G^2_0\left(p\right) 
\,G_0\left(q\right) 
\end{eqnarray}

Two cases need to be distinguished:

\begin{description}
\item[$p$ is small:]
Because of Eq.~(\ref{sig_1_ex_appr}), the first squared bracket delivers a
factor~$p^2$. Additional $p^2$ factors result from the third bracket and the
3D integration, respectively, so that we finally arrive at a $p^6$ factor.
From the identity $G_0^2\left(p,z\right)
\propto \frac{\partial}{\partial z}\,G_0\left(p,z\right)$, we obtain a
nonanalyticity~$\propto~z^{3/2}$. Note that uneven occurrences of ${\bf p}$ and/or
${\bf q}$, such as an
isolated product~${\bf pq}$, are not rotation invariant and therefore do not contribute
to the integral.
\item[$q$ is small:] 
Now the second bracket delivers an additional nonanalyticity \cite{comment3}
  $q^2$, which
again produces a $z^{3/2}$ nonanalyticity.
\end{description}

We therefore conclude for this group of diagrams

\begin{equation}
\lim_{z\to
    0}\,\lim_{k\to 0}\,{\rm
    Im}\left[\,\Sigma_\alpha^{(2)}\left(k,z\right)\,\right]\propto z^{3/2}k^2
\end{equation}

\subsection{$\Sigma_\beta^{(2)}\left(k,z\right)$: Irreducible Diagrams  $(1\,2 \cdots 2 \cdots 1)$}\label{sec_sig_beta}

Since sites~$1$ and $2$ are connected directly via $t^o_{12}$ or $t^d_{12}$,
the diagrams contain only three~$t$'s and two $G_0$'s. Therefore, only eight
diagrams are possible:

\subsubsection*{OOO:}

\begin{equation}\label{OOO_ex}
\parbox{4.5cm}{
\begin{fmfgraph*}(4.5,3) \fmfpen{thick}
\fmfsurround{z1,d1,z2,s1,d2}
\fmf{fermion}{s1,s2} 
\fmf{heavy,right}{s2,z1} 
\fmf{fermion,right}{z1,s2} 
\fmf{heavy}{s2,z2} 
\fmf{fermion}{z2,s1} 
\fmfdot{z1,z2,s2}
\fmfblob{4thick}{s1}
\fmflabel{{\bf 1}}{s1}
\fmflabel{{\bf 2}}{s2}
\end{fmfgraph*}}
\quad
=
\quad
\rho\int 
d{\bf p}\,d{\bf q}\,
\,t^2\left(p\right)
\,t\left(q\right)
\,G_0\left(p\right) 
\,G_0\left(q\right) 
\end{equation}

\subsubsection*{OOD:}

\begin{equation}\label{OOD_ex}
\parbox{4.5cm}{
\begin{fmfgraph*}(4.5,3) \fmfpen{thick}
\fmfsurround{z1,d1,z2,s1,d2}
\fmf{fermion}{s1,s2} 
\fmf{heavy,right}{s2,z1} 
\fmf{fermion,right}{z1,s2} 
\fmf{heavy}{s2,z2} 
\fmf{scalar}{z2,s1} 
\fmfdot{z1,s2}
\fmfblob{4thick}{s1,z2}
\fmflabel{{\bf 1}}{s1}
\fmflabel{{\bf 2}}{s2}
\end{fmfgraph*}}
\quad
=
\quad
-\,\rho\int 
d{\bf p}\,d{\bf q}\,
\,t\left({\bf k}-{\bf p}\right)
\,t\left(p\right)
\,t\left(q\right)
\,G_0\left(p\right) 
\,G_0\left(q\right) 
\end{equation}

\subsubsection*{ODO:}

\begin{equation}\label{ODO_ex}
\parbox{4.5cm}{
\begin{fmfgraph*}(4,3) \fmfpen{thick}
\fmfsurround{d1,z1,z2,s1,s2}
\fmf{fermion}{s1,s2} 
\fmf{heavy}{s2,z1} 
\fmf{scalar,left}{z1,s2}
\fmf{heavy}{z1,z2} 
\fmf{fermion}{z2,s1} 
\fmfdot{z1,z2,s2}
\fmfblob{4thick}{s1}
\fmflabel{{\bf 1}}{s1}
\fmflabel{{\bf 2}}{s2}
\end{fmfgraph*}}
\quad
=
\quad
-\,\rho\int 
d{\bf p}\,d{\bf q}\,
\,t^2\left(p\right)
\,t\left({\bf p} - {\bf q}\right)
\,G_0\left(p\right) 
\,G_0\left(q\right) 
\end{equation}

\subsubsection*{ODD:}

\begin{equation}\label{ODD_ex}
\parbox{4.5cm}{
\begin{fmfgraph*}(4,3) \fmfpen{thick}
\fmfsurround{d1,z1,z2,s1,s2}
\fmf{fermion}{s1,s2} 
\fmf{heavy}{s2,z1} 
\fmf{scalar,left}{z1,s2}
\fmf{heavy}{z1,z2} 
\fmf{scalar}{z2,s1} 
\fmfdot{z1,s2}
\fmfblob{4thick}{s1,z2}
\fmflabel{{\bf 1}}{s1}
\fmflabel{{\bf 2}}{s2}
\end{fmfgraph*}}
\quad
=
\quad
\rho\int 
d{\bf p}\,d{\bf q}\,
\,t\left({\bf k}-{\bf p}\right)
\,t\left(p\right)
\,t\left({\bf p} - {\bf q}\right)
\,G_0\left(p\right) 
\,G_0\left(q\right) 
\end{equation}

\subsubsection*{DOO:}

\begin{equation}\label{DOO_ex}
\parbox{4.5cm}{
\begin{fmfgraph*}(4,3) \fmfpen{thick}
\fmfsurround{s2,z2,s1,z1}
\fmf{scalar}{s1,s2} 
\fmf{heavy}{s1,z2} 
\fmf{fermion}{z2,s2}
\fmf{heavy}{s2,z1} 
\fmf{fermion}{z1,s1} 
\fmfdot{z1,z2,s2}
\fmfblob{4thick}{s1}
\fmflabel{{\bf 1}}{s1}
\fmflabel{{\bf 2}}{s2}
\end{fmfgraph*}}
\quad
=
\quad
-\,\rho\int 
d{\bf p}\,d{\bf q}\,
\,t\left(p\right)
\,t\left({\bf p} - {\bf q}\right)
\,t\left(q\right)
\,G_0\left(p\right) 
\,G_0\left(q\right) 
\end{equation}

\subsubsection*{DOD:}

\begin{equation}\label{DOD_ex}
\parbox{4.5cm}{
\begin{fmfgraph*}(4,3) \fmfpen{thick}
\fmfsurround{s2,z2,s1,z1}
\fmf{scalar}{s1,s2} 
\fmf{heavy}{s1,z2} 
\fmf{fermion}{z2,s2}
\fmf{heavy}{s2,z1} 
\fmf{scalar}{z1,s1} 
\fmfdot{z2,s2}
\fmfblob{4thick}{s1,z1}
\fmflabel{{\bf 1}}{s1}
\fmflabel{{\bf 2}}{s2}
\end{fmfgraph*}}
\quad
=
\quad
\rho\int 
d{\bf p}\,d{\bf q}\,
\,t\left({\bf k} - {\bf p}\right)
\,t\left({\bf p} - {\bf q}\right)
\,t\left(q\right)
\,G_0\left(p\right) 
\,G_0\left(q\right) 
\end{equation}

\subsubsection*{DDO:}

\begin{equation}\label{DDO_ex}
\parbox{4.5cm}{
\begin{fmfgraph*}(4,3) \fmfpen{thick}
\fmfsurround{s2,z1,z2,s1}
\fmf{scalar}{s1,s2} 
\fmf{heavy}{s1,z1} 
\fmf{scalar}{z1,s2}
\fmf{heavy}{z1,z2} 
\fmf{fermion}{z2,s1} 
\fmfdot{z1,z2}
\fmfblob{4thick}{s1}
\fmflabel{{\bf 1}}{s1}
\fmflabel{{\bf 2}}{s2}
\end{fmfgraph*}}
\quad
=
\quad
\rho\int 
d{\bf p}\,d{\bf q}\,
\,t\left(p\right)
\,t^2\left({\bf p} - {\bf q}\right)
\,G_0\left(p\right) 
\,G_0\left(q\right) 
\end{equation}

\subsubsection*{DDD:}

\begin{equation}\label{DDD_ex}
\parbox{4.5cm}{
\begin{fmfgraph*}(4,3) \fmfpen{thick}
\fmfsurround{s2,z1,z2,s1}
\fmf{scalar}{s1,s2} 
\fmf{heavy}{s1,z1} 
\fmf{scalar}{z1,s2}
\fmf{heavy}{z1,z2} 
\fmf{scalar}{z2,s1} 
\fmfdot{z1}
\fmfblob{4thick}{s1,z2}
\fmflabel{{\bf 1}}{s1}
\fmflabel{{\bf 2}}{s2}
\end{fmfgraph*}}
\quad
=
\quad
-\,\rho\int 
d{\bf p}\,d{\bf q}\,
\,t\left({\bf k} - {\bf p}\right)
\,t^2\left({\bf p} - {\bf q}\right)
\,G_0\left(p\right) 
\,G_0\left(q\right) 
\end{equation}

\subsubsection*{Added together:}

\begin{eqnarray}
\lefteqn{\Sigma_\beta^{(2)}\left(k,z\right) =}
\\\nonumber
&&
-\,\rho\int
d{\bf p}\,d{\bf q}\,
\left[\,t\left({\bf k} - {\bf p}\right) - t\left(p\right)\,\right]
\,\left[\,t\left({\bf p} - {\bf q}\right) - t\left(p\right)\,\right]
\,\left[\,t\left({\bf p} - {\bf q}\right) - t\left(q\right)\,\right]
\, G_0\left(p\right) 
\,G_0\left(q\right) 
\end{eqnarray}

Because of (\ref{tkpk}), the first bracket gives us only a $k^2$. Depending on
whether $p$ or $q$ are small, the third, respectively second, bracket delivers
the required additional $p^2$, respectively $q^2$, to obtain:

\begin{equation}
\lim_{z\to
    0}\,\lim_{k\to 0}\,{\rm
    Im}\left[\,\Sigma_\beta^{(2)}\left(k,z\right)\,\right]\propto z^{3/2}k^2
\end{equation}

Note a particular property of diagrams~${\bf DDO}$ and ${\bf DDD}$: 

If the propagator in the middle of the diagram collapses to the diagonal ${\cal G}_0$, as
explained above, these diagrams 
are factorizable at site~$1$ and therefore not irreducible anymore. To avoid double
counting, the diagonal term~${\cal G}_0$ must therefore be subtracted from
this propagator. 

It can be easily verified, however, that the $z^{3/2}$--nonanalyticity is not
affected by this.

\subsection{$\Sigma_\gamma^{(2)}\left(k,z\right)$: Irreducible Diagrams $(1 * 2 \cdots 1 \cdots 2)$}

The crossover topology of these 16 diagrams leads to more intricate
convolution integrals.

Based on the following abbreviations

\begin{equation}
\begin{array}{ccc}
\bfalpha &:=& {\bf k - p - q} \\
\bfbeta &:=& {\bf p} \\
\bfgamma &:=& {\bf q} \\
{\bf a} &:=& {\bf p + q} \\
{\bf b} &:=& {\bf k - p} \\
{\bf c} &:=& {\bf k - q}
\end{array}
\end{equation}

we obtain

\subsubsection*{OOOO:}

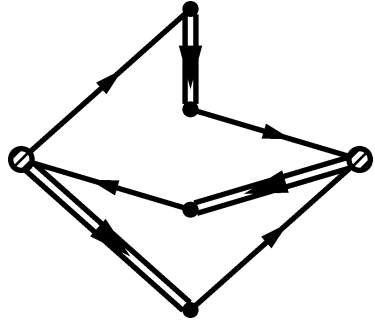
\begin{equation}\label{OOOO_cross}
\parbox{4.5cm}{
\begin{fmfgraph*}(4.5,4) \fmfpen{thick}
\fmfleft{s1}
\fmfright{s2}
\fmftop{z1}
\fmfbottom{z4}
\fmf{fermion}{s1,z1} 
\fmf{phantom}{z1,s2} 
\fmf{heavy}{z1,z2} 
\fmf{fermion}{z2,s2} 
\fmf{phantom}{s1,z2} 
\fmf{heavy}{s2,z3} 
\fmf{fermion}{z3,s1}
\fmf{heavy}{s1,z4} 
\fmf{fermion}{z4,s2}
\fmf{phantom}{z3,z4} 
\fmfdot{z1,z2,z3,z4}
\fmfblob{4thick}{s1,s2}
\fmflabel{{\bf 1}}{s1}
\fmflabel{{\bf 2}}{s2}
\end{fmfgraph*}}
\end{equation}

\begin{displaymath}
\rho^2\,\int\int d\bfbeta\,d\bfgamma\quad
G_0\left(\bfalpha\right)G_0\left(\bfbeta\right)G_0\left(\bfgamma\right)
\,\cdot\,
t\left(\bfalpha\right)t\left(\bfbeta\right)
t^2\left(\bfgamma\right)
\end{displaymath}

\subsubsection*{OOOD:}

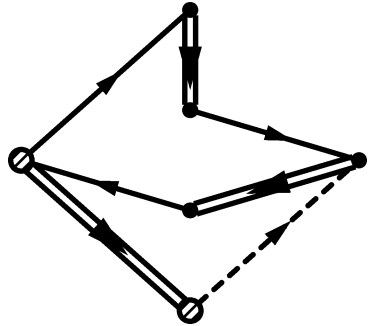
\begin{equation}\label{OOOD_cross}
\parbox{4.5cm}{
\begin{fmfgraph*}(4.5,4) \fmfpen{thick}
\fmfleft{s1}
\fmfright{s2}
\fmftop{z1}
\fmfbottom{z4}
\fmf{fermion}{s1,z1} 
\fmf{phantom}{z1,s2} 
\fmf{heavy}{z1,z2} 
\fmf{fermion}{z2,s2} 
\fmf{phantom}{s1,z2} 
\fmf{heavy}{s2,z3} 
\fmf{fermion}{z3,s1}
\fmf{heavy}{s1,z4} 
\fmf{scalar}{z4,s2}
\fmf{phantom}{z3,z4} 
\fmfdot{z1,z2,z3,s2}
\fmfblob{4thick}{s1,z4}
\fmflabel{{\bf 1}}{s1}
\fmflabel{{\bf 2}}{s2}
\end{fmfgraph*}}
\end{equation}

\begin{displaymath}
-\,
\rho^2\,\int\int d\bfbeta\,d\bfgamma\quad
G_0\left(\bfalpha\right)G_0\left(\bfbeta\right)G_0\left(\bfgamma\right)
\,\cdot\,
t\left({\bf a}\right) t\left(\bfbeta\right) t^2\left(\bfgamma\right)
\end{displaymath}

\subsubsection*{OODO:}

\begin{equation}\label{OODO_cross}
\parbox{4.5cm}{
\begin{fmfgraph*}(4.5,4) \fmfpen{thick}
\fmfleft{s1}
\fmfright{s2}
\fmftop{z1}
\fmfbottom{z4}
\fmf{fermion}{s1,z1} 
\fmf{phantom}{z1,s2} 
\fmf{heavy}{z1,z2} 
\fmf{fermion}{z2,s2} 
\fmf{phantom}{s1,z2} 
\fmf{heavy}{s2,z3} 
\fmf{scalar}{z3,s1}
\fmf{heavy}{z3,z4} 
\fmf{fermion}{z4,s2}
\fmf{phantom}{s1,z4} 
\fmfdot{z1,z2,z3,z4}
\fmfblob{4thick}{s1,s2}
\fmflabel{{\bf 1}}{s1}
\fmflabel{{\bf 2}}{s2}
\end{fmfgraph*}}
\end{equation}

\begin{displaymath}
-\,
\rho^2\,\int\int d\bfbeta\,d\bfgamma\quad
G_0\left(\bfalpha\right)G_0\left(\bfbeta\right)G_0\left(\bfgamma\right)
\,\cdot\,
t^2\left(\bfalpha\right) t\left({\bf a}\right) t\left(\bfgamma\right)
\end{displaymath}

\subsubsection*{OODD:}

\begin{equation}\label{OODD_cross}
\parbox{4.5cm}{
\begin{fmfgraph*}(4.5,4) \fmfpen{thick}
\fmfleft{s1}
\fmfright{s2}
\fmftop{z1}
\fmfbottom{z4}
\fmf{fermion}{s1,z1} 
\fmf{phantom}{z1,s2} 
\fmf{heavy}{z1,z2} 
\fmf{fermion}{z2,s2} 
\fmf{phantom}{s1,z2} 
\fmf{heavy}{s2,z3} 
\fmf{scalar}{z3,s1}
\fmf{heavy}{z3,z4} 
\fmf{scalar}{z4,s2}
\fmf{phantom}{s1,z4} 
\fmfdot{z1,z2,z3,s2}
\fmfblob{4thick}{s1,z4}
\fmflabel{{\bf 1}}{s1}
\fmflabel{{\bf 2}}{s2}
\end{fmfgraph*}}
\end{equation}

\begin{displaymath}
\rho^2\,\int\int d\bfbeta\,d\bfgamma\quad
G_0\left(\bfalpha\right)G_0\left(\bfbeta\right)G_0\left(\bfgamma\right)
\,\cdot\,
t\left({\bf b}\right) t\left({\bf c}\right) t^2\left(\bfbeta\right)
\end{displaymath}

\subsubsection*{ODOO:}

\begin{equation}\label{ODOO_cross}
\parbox{4.5cm}{
\begin{fmfgraph*}(4.5,4) \fmfpen{thick}
\fmfleft{s1}
\fmfright{s2}
\fmftop{z1}
\fmfbottom{z4}
\fmf{fermion}{s1,z1} 
\fmf{phantom}{z1,s2} 
\fmf{heavy}{z1,z2} 
\fmf{scalar}{z2,s2} 
\fmf{phantom}{s1,z2} 
\fmf{heavy,tension=0}{z2,z3} 
\fmf{fermion}{z3,s1}
\fmf{phantom}{z3,s2}
\fmf{heavy}{s1,z4} 
\fmf{fermion}{z4,s2}
\fmf{phantom}{z3,z4} 
\fmfdot{z1,z2,z3,z4}
\fmfblob{4thick}{s1,s2}
\fmflabel{{\bf 1}}{s1}
\fmflabel{{\bf 2}}{s2}
\end{fmfgraph*}}
\end{equation}

\begin{displaymath}
-\,\rho^2\,\int\int d\bfbeta\,d\bfgamma\quad
G_0\left(\bfalpha\right)G_0\left(\bfbeta\right)G_0\left(\bfgamma\right)
\,\cdot\,
t\left(\bfalpha\right) t\left({\bf a}\right) t\left(\bfbeta\right) t\left(\bfgamma\right)
\end{displaymath}

\subsubsection*{ODOD:}

\begin{equation}\label{ODOD_cross}
\parbox{4.5cm}{
\begin{fmfgraph*}(4.5,4) \fmfpen{thick}
\fmfleft{s1}
\fmfright{s2}
\fmftop{z1}
\fmfbottom{z4}
\fmf{fermion}{s1,z1} 
\fmf{phantom}{z1,s2} 
\fmf{heavy}{z1,z2} 
\fmf{scalar}{z2,s2} 
\fmf{phantom}{s1,z2} 
\fmf{heavy,tension=0}{z2,z3} 
\fmf{fermion}{z3,s1}
\fmf{phantom}{z3,s2}
\fmf{heavy}{s1,z4} 
\fmf{scalar}{z4,s2}
\fmf{phantom}{z3,z4} 
\fmfdot{z1,z2,z3}
\fmfblob{4thick}{s1,z4}
\fmflabel{{\bf 1}}{s1}
\fmflabel{{\bf 2}}{s2}
\end{fmfgraph*}}
\end{equation}

\begin{displaymath}
\rho^2\,\int\int d\bfbeta\,d\bfgamma\quad
G_0\left(\bfalpha\right)G_0\left(\bfbeta\right)G_0\left(\bfgamma\right)
\,\cdot\,
t^2\left({\bf a}\right) t\left(\bfbeta\right) t\left(\bfgamma\right)
\end{displaymath}

\subsubsection*{ODDO:}

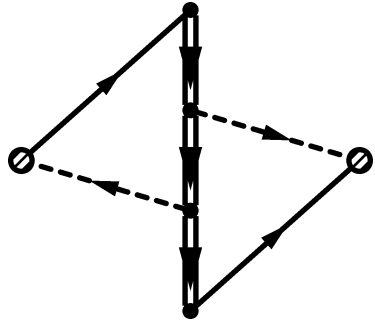
\begin{equation}\label{ODDO_cross}
\parbox{4.5cm}{
\begin{fmfgraph*}(4.5,4) \fmfpen{thick}
\fmfleft{s1}
\fmfright{s2}
\fmftop{z1}
\fmfbottom{z4}
\fmf{fermion}{s1,z1} 
\fmf{phantom}{z1,s2} 
\fmf{scalar}{z2,s2} 
\fmf{phantom}{s1,z2} 
\fmf{heavy}{z1,z2} 
\fmf{heavy,tension=0}{z2,z3} 
\fmf{heavy}{z3,z4} 
\fmf{scalar}{z3,s1}
\fmf{phantom}{z3,s2}
\fmf{fermion}{z4,s2}
\fmf{phantom}{s1,z4} 
\fmfdot{z1,z2,z3,z4}
\fmfblob{4thick}{s1,s2}
\fmflabel{{\bf 1}}{s1}
\fmflabel{{\bf 2}}{s2}
\end{fmfgraph*}}
\end{equation}

\begin{displaymath}
\rho^2\,\int\int d\bfbeta\,d\bfgamma\quad
G_0\left(\bfalpha\right)G_0\left(\bfbeta\right)G_0\left(\bfgamma\right)
\,\cdot\,
t\left({\bf b}\right) t\left({\bf c}\right) t\left(\bfbeta\right) t\left(\bfgamma\right)
\end{displaymath}

\subsubsection*{ODDD:}

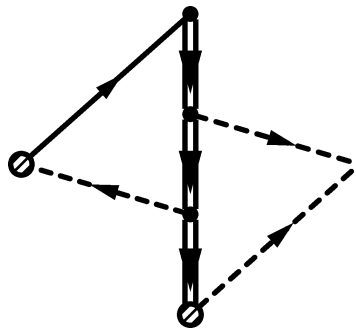
\begin{equation}\label{ODDD_cross}
\parbox{4.5cm}{
\begin{fmfgraph*}(4.5,4) \fmfpen{thick}
\fmfleft{s1}
\fmfright{s2}
\fmftop{z1}
\fmfbottom{z4}
\fmf{fermion}{s1,z1} 
\fmf{phantom}{z1,s2} 
\fmf{scalar}{z2,s2} 
\fmf{phantom}{s1,z2} 
\fmf{heavy}{z1,z2} 
\fmf{heavy,tension=0}{z2,z3} 
\fmf{heavy}{z3,z4} 
\fmf{scalar}{z3,s1}
\fmf{phantom}{z3,s2}
\fmf{scalar}{z4,s2}
\fmf{phantom}{s1,z4} 
\fmfdot{z1,z2,z3}
\fmfblob{4thick}{s1,z4}
\fmflabel{{\bf 1}}{s1}
\fmflabel{{\bf 2}}{s2}
\end{fmfgraph*}}
\end{equation}

\begin{displaymath}
-\,\rho^2\,\int\int d\bfbeta\,d\bfgamma\quad
G_0\left(\bfalpha\right)G_0\left(\bfbeta\right)G_0\left(\bfgamma\right)
t\left({\bf b}\right) t^2\left({\bf c}\right) t\left(\bfbeta\right)
\end{displaymath}

\subsubsection*{DOOO:}

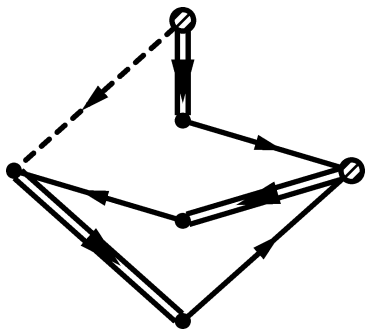
\begin{equation}\label{DOOO_cross}
\parbox{4.5cm}{
\begin{fmfgraph*}(4.5,4) \fmfpen{thick}
\fmfleft{s1}
\fmfright{s2}
\fmftop{z1}
\fmfbottom{z4}
\fmf{scalar}{z1,s1} 
\fmf{phantom}{z1,s2} 
\fmf{heavy}{z1,z2} 
\fmf{fermion}{z2,s2} 
\fmf{phantom}{s1,z2} 
\fmf{heavy}{s2,z3} 
\fmf{fermion}{z3,s1}
\fmf{heavy}{s1,z4} 
\fmf{fermion}{z4,s2}
\fmf{phantom}{z3,z4} 
\fmfdot{s1,z2,z3,z4}
\fmfblob{4thick}{z1,s2}
\fmflabel{{\bf 1}}{s1}
\fmflabel{{\bf 2}}{s2}
\end{fmfgraph*}}
\end{equation}

\begin{displaymath}
-\,\rho^2\,\int\int d\bfbeta\,d\bfgamma\quad
G_0\left(\bfalpha\right)G_0\left(\bfbeta\right)G_0\left(\bfgamma\right)
\,\cdot\,
t\left(\bfalpha\right) t\left({\bf a}\right) t\left(\bfbeta\right) t\left(\bfgamma\right)
\end{displaymath}

\subsubsection*{DOOD:}

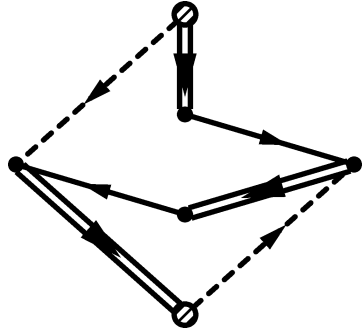
\begin{equation}\label{DOOD_cross}
\parbox{4.5cm}{
\begin{fmfgraph*}(4.5,4) \fmfpen{thick}
\fmfleft{s1}
\fmfright{s2}
\fmftop{z1}
\fmfbottom{z4}
\fmf{scalar}{z1,s1} 
\fmf{phantom}{z1,s2} 
\fmf{heavy}{z1,z2} 
\fmf{fermion}{z2,s2} 
\fmf{phantom}{s1,z2} 
\fmf{heavy}{s2,z3} 
\fmf{fermion}{z3,s1}
\fmf{heavy}{s1,z4} 
\fmf{scalar}{z4,s2}
\fmf{phantom}{z3,z4} 
\fmfdot{s1,z2,z3,s2}
\fmfblob{4thick}{z1,z4}
\fmflabel{{\bf 1}}{s1}
\fmflabel{{\bf 2}}{s2}
\end{fmfgraph*}}
\end{equation}

\begin{displaymath}
\rho^2\,\int\int d\bfbeta\,d\bfgamma\quad
G_0\left(\bfalpha\right)G_0\left(\bfbeta\right)G_0\left(\bfgamma\right)
\,\cdot\,
t\left(\bfalpha\right) t\left({\bf b}\right) t\left({\bf c}\right) t\left(\bfbeta\right)
\end{displaymath}

\subsubsection*{DODO:}

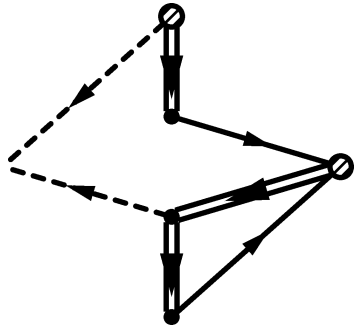
\begin{equation}\label{DODO_cross}
\parbox{4.5cm}{
\begin{fmfgraph*}(4.5,4) \fmfpen{thick}
\fmfleft{s1}
\fmfright{s2}
\fmftop{z1}
\fmfbottom{z4}
\fmf{scalar}{z1,s1} 
\fmf{phantom}{z1,s2} 
\fmf{heavy}{z1,z2} 
\fmf{fermion}{z2,s2} 
\fmf{phantom}{s1,z2} 
\fmf{heavy}{s2,z3} 
\fmf{scalar}{z3,s1}
\fmf{heavy}{z3,z4} 
\fmf{fermion}{z4,s2}
\fmf{phantom}{s1,z4} 
\fmfdot{z2,z3,z4}
\fmfblob{4thick}{z1,s2}
\fmflabel{{\bf 1}}{s1}
\fmflabel{{\bf 2}}{s2}
\end{fmfgraph*}}
\end{equation}

\begin{displaymath}
\rho^2\,\int\int d\bfbeta\,d\bfgamma\quad
G_0\left(\bfalpha\right)G_0\left(\bfbeta\right)G_0\left(\bfgamma\right)
\,\cdot\,
t\left(\bfalpha\right) t^2\left({\bf a}\right) t\left(\bfgamma\right)
\end{displaymath}

\subsubsection*{DODD:}

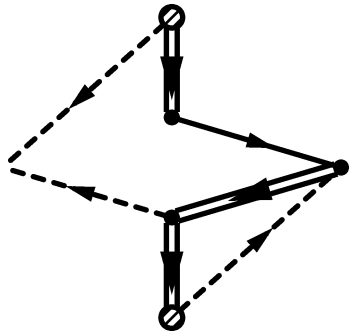
\begin{equation}\label{DODD_cross}
\parbox{4.5cm}{
\begin{fmfgraph*}(4.5,4) \fmfpen{thick}
\fmfleft{s1}
\fmfright{s2}
\fmftop{z1}
\fmfbottom{z4}
\fmf{scalar}{z1,s1} 
\fmf{phantom}{z1,s2} 
\fmf{heavy}{z1,z2} 
\fmf{fermion}{z2,s2} 
\fmf{phantom}{s1,z2} 
\fmf{heavy}{s2,z3} 
\fmf{scalar}{z3,s1}
\fmf{heavy}{z3,z4} 
\fmf{scalar}{z4,s2}
\fmf{phantom}{s1,z4} 
\fmfdot{z2,z3,s2}
\fmfblob{4thick}{z1,z4}
\fmflabel{{\bf 1}}{s1}
\fmflabel{{\bf 2}}{s2}
\end{fmfgraph*}}
\end{equation}

\begin{displaymath}
-\,\rho^2\,\int\int d\bfbeta\,d\bfgamma\quad
G_0\left(\bfalpha\right)G_0\left(\bfbeta\right)G_0\left(\bfgamma\right)
\,\cdot\,
t^2\left({\bf b}\right) t\left({\bf c}\right) t\left(\bfbeta\right)
\end{displaymath}

\subsubsection*{DDOO:}

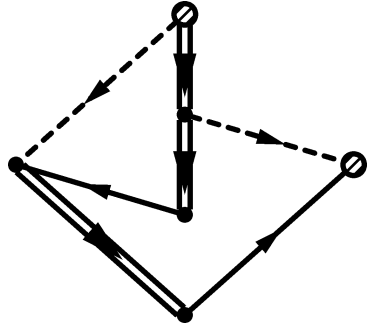
\begin{equation}\label{DDOO_cross}
\parbox{4.5cm}{
\begin{fmfgraph*}(4.5,4) \fmfpen{thick}
\fmfleft{s1}
\fmfright{s2}
\fmftop{z1}
\fmfbottom{z4}
\fmf{scalar}{z1,s1} 
\fmf{phantom}{z1,s2} 
\fmf{heavy}{z1,z2} 
\fmf{scalar}{z2,s2} 
\fmf{phantom}{s1,z2} 
\fmf{heavy,tension=0}{z2,z3} 
\fmf{fermion}{z3,s1}
\fmf{phantom}{z3,s2}
\fmf{heavy}{s1,z4} 
\fmf{fermion}{z4,s2}
\fmf{phantom}{z3,z4} 
\fmfdot{s1,z2,z3,z4}
\fmfblob{4thick}{z1,s2}
\fmflabel{{\bf 1}}{s1}
\fmflabel{{\bf 2}}{s2}
\end{fmfgraph*}}
\end{equation}

\begin{displaymath}
\rho^2\,\int\int d\bfbeta\,d\bfgamma\quad
G_0\left(\bfalpha\right)G_0\left(\bfbeta\right)G_0\left(\bfgamma\right)
\,\cdot\,
t\left(\bfalpha\right) t\left({\bf b}\right) t\left({\bf c}\right) t\left(\bfgamma\right)
\end{displaymath}

\subsubsection*{DDOD:}

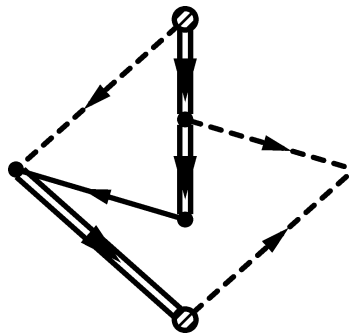
\begin{equation}\label{DDOD_cross}
\parbox{4.5cm}{
\begin{fmfgraph*}(4.5,4) \fmfpen{thick}
\fmfleft{s1}
\fmfright{s2}
\fmftop{z1}
\fmfbottom{z4}
\fmf{scalar}{z1,s1} 
\fmf{phantom}{z1,s2} 
\fmf{heavy}{z1,z2} 
\fmf{scalar}{z2,s2} 
\fmf{phantom}{s1,z2} 
\fmf{heavy,tension=0}{z2,z3} 
\fmf{fermion}{z3,s1}
\fmf{phantom}{z3,s2}
\fmf{heavy}{s1,z4} 
\fmf{scalar}{z4,s2}
\fmf{phantom}{z3,z4} 
\fmfdot{s1,z2,z3}
\fmfblob{4thick}{z1,z4}
\fmflabel{{\bf 1}}{s1}
\fmflabel{{\bf 2}}{s2}
\end{fmfgraph*}}
\end{equation}

\begin{displaymath}
-\,\rho^2\,\int\int d\bfbeta\,d\bfgamma\quad
G_0\left(\bfalpha\right)G_0\left(\bfbeta\right)G_0\left(\bfgamma\right)
\,\cdot\,
t\left(\bfalpha\right) t\left({\bf b}\right) t^2\left({\bf c}\right)
\end{displaymath}

\subsubsection*{DDDO:}

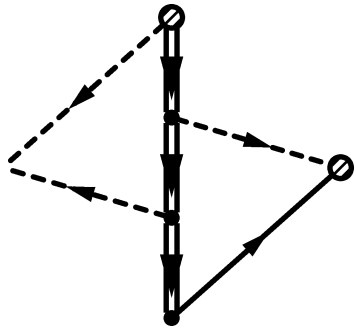
\begin{equation}\label{DDDO_cross}
\parbox{4.5cm}{
\begin{fmfgraph*}(4.5,4) \fmfpen{thick}
\fmfleft{s1}
\fmfright{s2}
\fmftop{z1}
\fmfbottom{z4}
\fmf{scalar}{z1,s1} 
\fmf{phantom}{z1,s2} 
\fmf{scalar}{z2,s2} 
\fmf{phantom}{s1,z2} 
\fmf{heavy}{z1,z2} 
\fmf{heavy,tension=0}{z2,z3} 
\fmf{heavy}{z3,z4} 
\fmf{scalar}{z3,s1}
\fmf{phantom}{z3,s2}
\fmf{fermion}{z4,s2}
\fmf{phantom}{s1,z4} 
\fmfdot{z2,z3,z4}
\fmfblob{4thick}{z1,s2}
\fmflabel{{\bf 1}}{s1}
\fmflabel{{\bf 2}}{s2}
\end{fmfgraph*}}
\end{equation}

\begin{displaymath}
-\,\rho^2\,\int\int d\bfbeta\,d\bfgamma\quad
G_0\left(\bfalpha\right)G_0\left(\bfbeta\right)G_0\left(\bfgamma\right)
\,\cdot\,
t^2\left({\bf b}\right) t\left({\bf c}\right) t\left(\bfgamma\right)
\end{displaymath}

\subsubsection*{DDDD:}

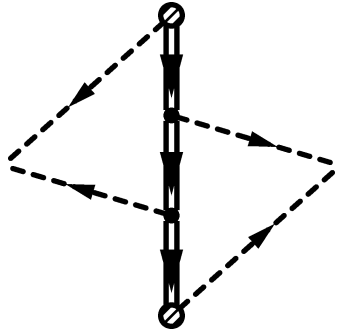
\begin{equation}\label{DDDD_cross}
\parbox{4.5cm}{
\begin{fmfgraph*}(4.5,4) \fmfpen{thick}
\fmfleft{s1}
\fmfright{s2}
\fmftop{z1}
\fmfbottom{z4}
\fmf{scalar}{z1,s1} 
\fmf{phantom}{z1,s2} 
\fmf{scalar}{z2,s2} 
\fmf{phantom}{s1,z2} 
\fmf{heavy}{z1,z2} 
\fmf{heavy,tension=0}{z2,z3} 
\fmf{heavy}{z3,z4} 
\fmf{scalar}{z3,s1}
\fmf{phantom}{z3,s2}
\fmf{scalar}{z4,s2}
\fmf{phantom}{s1,z4} 
\fmfdot{z2,z3}
\fmfblob{4thick}{z1,z4}
\fmflabel{{\bf 1}}{s1}
\fmflabel{{\bf 2}}{s2}
\end{fmfgraph*}}
\end{equation}

\begin{displaymath}
\rho^2\,\int\int d\bfbeta\,d\bfgamma\quad
G_0\left(\bfalpha\right)G_0\left(\bfbeta\right)G_0\left(\bfgamma\right)
\,\cdot\,
t^2\left({\bf b}\right) t^2\left({\bf c}\right)
\end{displaymath}

\subsubsection*{Collecting Terms}

To derive a usable expression for the sum of these 16 diagrams, we have to
exploit the symmetries of the problem.

Transforming variables ${\bf p},\,{\bf q} \to {\bf
  \widetilde{p}},\,{\bf \widetilde{q}}$ allows to arbitrarily permute
$\bfalpha$, $\bfbeta$, $\bfgamma$ under the boundary condition that 
${\bf a}$, ${\bf b}$, ${\bf c}$ perform the same permutation, as indicated in
the following table:

\begin{equation}
\begin{array}{cccccc}
\bfalpha & \bfbeta & \bfgamma & {\bf a} & {\bf b} & {\bf c} \\
\bfalpha & \bfgamma & \bfbeta & {\bf a} & {\bf c} & {\bf b} \\
\bfbeta & \bfalpha & \bfgamma & {\bf b} & {\bf a} & {\bf c} \\
\bfbeta & \bfgamma & \bfalpha & {\bf b} & {\bf c} & {\bf a} \\
\bfgamma & \bfalpha & \bfbeta & {\bf c} & {\bf a} & {\bf b} \\
\bfgamma & \bfbeta & \bfalpha & {\bf c} & {\bf b} & {\bf a}
\end{array}
\end{equation}

For example, the transformation

\begin{equation}
{\bf \widetilde{p}} := {\bf k - p - q} \qquad\mbox{und}\qquad
{\bf \widetilde{q}} := {\bf p}
\end{equation}

leads to

\begin{eqnarray}
{\bf k - p - q} &\to& {\bf \widetilde{p}}\\
{\bf p} &\to& {\bf \widetilde{q}}\\
{\bf q} &\to& {\bf k - \widetilde{p} - \widetilde{q}} \\
{\bf p + q} &\to& {\bf k - \widetilde{p}}\\
{\bf k - p} &\to& {\bf k - \widetilde{q}}\\
{\bf k - q} &\to& {\bf \widetilde{p} + \widetilde{q}}
\end{eqnarray}

and thus the permutation~$\bfalpha\,\bfbeta\,\bfgamma\,{\bf a\,b\,c} \to
\bfbeta\,\bfgamma\,\bfalpha\,{\bf b\,c\,a}$.

The product~$G_0\left(\bfalpha\right)G_0\left(\bfbeta\right)G_0\left(\bfgamma\right)$
is invariant with respect to these permutations and because of

\begin{equation}
\int\int d\bfalpha\,d\bfbeta = \int\int d\bfalpha\,d\bfgamma = \int\int
d\bfbeta\,d\bfgamma
\end{equation}

the integration variables can be chosen freely. After suitably regrouping the
$t$--factors, we obtain for the sum of all diagrams the expression

\begin{eqnarray}\nonumber
K &:=& {\bf OOOO} + \cdots + {\bf 
  DDDD}
\\\nonumber
&=& \rho^2\,\int\int
d\bfbeta\,d\bfgamma\quad
G_0\left(\bfalpha\right)G_0\left(\bfbeta\right)G_0\left(\bfgamma\right)\quad\times
\\\nonumber
&\times&
\left[\,t\left(\bfbeta\right) - t\left({\bf b}\right)\,\right]
\cdot
\left[\,t\left(\bfgamma\right) - t\left({\bf c}\right)\,\right]\cdot
\left[\,t\left(\bfgamma\right) - t\left({\bfbeta}\right)\,\right]
\cdot
\left[\,t\left(\bfalpha\right) + t\left({\bfbeta}\right)\,\right]
\end{eqnarray}

Since we integrate over two variables only, we have to eliminate one 
$G_0$--factor. To this end, we apply a partial fraction decomposition

\begin{equation}
\rho\left[\,t\left(\bfgamma\right) - t\left({\bfbeta}\right)\,\right]\cdot
G_0\left(\bfbeta\right)G_0\left(\bfgamma\right) = 
\left[\,G_0\left(\bfgamma\right) - G_0\left(\bfbeta\right)\,\right]
\end{equation}

Some further permutations and regrouping lead to

\begin{eqnarray}\nonumber
K &=& \rho^2\,\int\int
d\bfbeta\,d\bfgamma\quad
G_0\left(\bfbeta\right)G_0\left(\bfgamma\right)\quad\times
\\\nonumber
&\times&
\left[\,t\left(\bfalpha\right) - t\left({\bf a}\right)\,\right]
\cdot
\left[\,t\left(\bfalpha\right) - t\left({\bfgamma}\right)\,\right]\cdot
\left[\,t\left(\bfgamma\right) - t\left({\bf c}\right)\,\right]
\end{eqnarray}

After reinserting the above definitions, we can expand for small~$k$:

\begin{eqnarray}\nonumber
t\left(\bfalpha\right) - t\left({\bf a}\right) &=&
t\left({\bf k - p - q}\right) - t\left({\bf p + q}\right)
\\\nonumber
&=& 
\frac{t^\prime\left({\bf p + q}\right)}{\left|{\bf p + q}\right|}\cdot
{\bf k}\left({\bf p + q}\right) + {\cal O}\left(k^2\right)
\end{eqnarray}

\begin{eqnarray}\nonumber
t\left({\bf c}\right) - t\left(\bfgamma\right) &=&
t\left({\bf k - q}\right) - t\left(q\right)
\\\nonumber
&=& \frac{t^\prime\left(q\right)}{q}\cdot
{\bf kq} + {\cal O}\left(k^2\right)
\end{eqnarray}

We thus obtain $\lim_{k\to 0}K \propto k^2$, as required. Since we only
consider the lowest-order term, we may use

\begin{equation}
t\left(\bfalpha\right) - t\left({\bfgamma}\right) =
t\left({\bf a}\right) - t\left({\bfgamma}\right) + {\cal O}\left(k\right)
\end{equation}

and set

\begin{equation}
t\left(\bfalpha\right) - t\left({\bfgamma}\right) \approx
t\left({\bf p + q}\right) - t\left(q\right)
\end{equation}

We thus finally arrive at

\begin{eqnarray}\label{K}
K &=& \rho^2\,\int\int
d{\bf p}\,d{\bf q}\quad
G_0\left({\bf p}\right)G_0\left({\bf q}\right)\quad\times
\\\nonumber
&\times&
\underbrace{\left[\,\frac{t^\prime\left({\bf p + q}\right)}{\left|{\bf p + q}\right|}\cdot
{\bf k}\left({\bf p + q}\right)\,\right]}_A
\cdot
\underbrace{\left[\,t\left({\bf p + q}\right) - t\left(q\right)\,\right]}_B
\cdot
\underbrace{\left[\,\frac{t^\prime\left(q\right)}{q}\cdot
{\bf kq}\,\right]}_C + {\cal O}\left(k^3\right)
\end{eqnarray}

For the analytical properties, we again have to consider two cases:

\subsubsection*{$q$ is small:}

Bracket~$C$ gives us a factor~$q$, but brackets~$A$ and $B$ approach a finite
value for $q\to 0$.

{\bf But:} After setting ${\bf q} = {\bf 0}$ in $A$ and $B$, both the integral
over ${\bf p}$ (because of the~${\bf
  kp}$ term in~$A$) and over ${\bf q}$ (due to the ${\bf kq}$ in
$C$) vanish due to lack of rotational invariance. Consequently, we have to
expand the fraction in $A$ to first order in $q$, which provides us with an additional
factor~${\bf pq} \propto q$. Now rotational invariance is preserved and a
$z^{3/2}$ nonanalyticity is obtained.

\subsubsection*{$p$ is small:}

Bracket $C$ remains finite and $B$ leads to a factor~${\bf pq}\propto p$. 

Setting ${\bf p = 0}$ in $A$ leads to the following structure
\begin{equation}
\int\int
d{\bf p}\,d{\bf q}\quad f\left(q\right)\cdot \left[\,{\bf
  pq}\,\right]\cdot\left[\,{\bf kq}\,\right]^2
\end{equation}

In order to restore rotational invariance, we again have to expand in $A$ to
first order in $p$, which yields an 
additional~${\bf pq}\propto p$ and leads to a $z^{3/2}$ nonanalyticity.

We thus conclude that

\begin{equation}
\lim_{z\to
    0}\,\lim_{k\to 0}\,{\rm
    Im}\left[\,\Sigma_\gamma^{(2)}\left(k,z\right)\,\right]\propto z^{3/2}k^2
\end{equation}

holds.

\subsection{$\Sigma_\delta^{(2)}\left(k,z\right)$: Irreducible Diagrams $(1\,2 \cdots 1 \cdots 2)$}

Similar to $\Sigma_\beta$ above, 8 diagrams need to be considered:

\subsubsection*{OOO:}

\begin{equation}\label{OOO_kr}
\parbox{4.5cm}{
\begin{fmfgraph*}(4,3) \fmfpen{thick}
\fmfsurround{s2,z2,s1,z1}
\fmf{fermion}{s1,s2} 
\fmf{heavy}{s1,z2} 
\fmf{fermion}{z2,s2}
\fmf{heavy}{s2,z1} 
\fmf{fermion}{z1,s1} 
\fmfdot{z1,z2}
\fmfblob{4thick}{s1,s2}
\fmflabel{{\bf 1}}{s1}
\fmflabel{{\bf 2}}{s2}
\end{fmfgraph*}}
\quad
=
\quad
\rho\,\int 
d{\bf p}\,d{\bf q}\,
\,t\left({\bf k} - {\bf p} + {\bf q}\right)
\,t\left(p\right)
\,t\left(q\right)
\,G_0\left(p\right) 
\,G_0\left(q\right) 
\end{equation}

\subsubsection*{OOD:}

\begin{equation}\label{OOD_kr}
\parbox{4.5cm}{
\begin{fmfgraph*}(4,3) \fmfpen{thick}
\fmfsurround{s2,z2,s1,z1}
\fmf{fermion}{s1,s2} 
\fmf{heavy}{s1,z2} 
\fmf{scalar}{z2,s2}
\fmf{heavy}{s2,z1} 
\fmf{fermion}{z1,s1} 
\fmfdot{z1,s2}
\fmfblob{4thick}{s1,z2}
\fmflabel{{\bf 1}}{s1}
\fmflabel{{\bf 2}}{s2}
\end{fmfgraph*}}
\quad
=
\quad
-\rho\,\int 
d{\bf p}\,d{\bf q}\,
\,t\left({\bf k} - {\bf p} + {\bf q}\right)
\,t\left({\bf k} - {\bf p}\right)
\,t\left(q\right)
\,G_0\left(p\right) 
\,G_0\left(q\right) 
\end{equation}

\subsubsection*{ODO:}

\begin{equation}\label{ODO_kr}
\parbox{4.5cm}{
\begin{fmfgraph*}(4,3) \fmfpen{thick}
\fmfsurround{s2,z1,z2,s1}
\fmf{fermion}{s2,s1} 
\fmf{heavy}{s1,z1} 
\fmf{scalar}{z1,s2}
\fmf{heavy}{z1,z2} 
\fmf{fermion}{z2,s1} 
\fmfdot{z1,z2}
\fmfblob{4thick}{s2,s1}
\fmflabel{{\bf 1}}{s2}
\fmflabel{{\bf 2}}{s1}
\end{fmfgraph*}}
\quad
=
\quad
-\rho\,\int 
d{\bf p}\,d{\bf q}\,
\,t\left({\bf k} - {\bf p} + {\bf q}\right)
\,t\left(p\right)
\,t\left({\bf p}-{\bf q}\right)
\,G_0\left(p\right) 
\,G_0\left(q\right) 
\end{equation}

\subsubsection*{ODD:}

\begin{equation}\label{ODD_kr}
\parbox{4.5cm}{
\begin{fmfgraph*}(4,3) \fmfpen{thick}
\fmfsurround{s2,z1,z2,s1}
\fmf{fermion}{s2,s1} 
\fmf{heavy}{s1,z1} 
\fmf{scalar}{z1,s2}
\fmf{heavy}{z1,z2} 
\fmf{scalar}{z2,s1} 
\fmfdot{z1,s1}
\fmfblob{4thick}{s2,z2}
\fmflabel{{\bf 1}}{s2}
\fmflabel{{\bf 2}}{s1}
\end{fmfgraph*}}
\quad
=
\quad
\rho\,\int 
d{\bf p}\,d{\bf q}\,
\,t\left({\bf k} - {\bf p} + {\bf q}\right)
\,t\left({\bf k} - {\bf p}\right)
\,t\left({\bf p}-{\bf q}\right)
\,G_0\left(p\right) 
\,G_0\left(q\right) 
\end{equation}

\subsubsection*{DOO:}

\begin{equation}\label{DOO_kr}
\parbox{4.5cm}{
\begin{fmfgraph*}(4.5,3) \fmfpen{thick}
\fmfsurround{z1,d1,z2,s1,d2}
\fmf{scalar}{s2,s1} 
\fmf{heavy,right}{s2,z1} 
\fmf{fermion,right}{z1,s2} 
\fmf{heavy}{s2,z2} 
\fmf{fermion}{z2,s1} 
\fmfdot{z1,z2}
\fmfblob{4thick}{s1,s2}
\fmflabel{{\bf 1}}{s2}
\fmflabel{{\bf 2}}{s1}
\end{fmfgraph*}}
\quad
=
\quad
-\rho\,\int 
d{\bf p}\,d{\bf q}\,
\,t\left({\bf k}-{\bf p}\right)
\,t\left(p\right)
\,t\left(q\right)
\,G_0\left(p\right) 
\,G_0\left(q\right) 
\end{equation}

\subsubsection*{DOD:}

\begin{equation}\label{DOD_kr}
\parbox{4.5cm}{
\begin{fmfgraph*}(4.5,3) \fmfpen{thick}
\fmfsurround{z1,d1,z2,s1,d2}
\fmf{scalar}{s2,s1} 
\fmf{heavy,right}{s2,z1} 
\fmf{fermion,right}{z1,s2} 
\fmf{heavy}{s2,z2} 
\fmf{scalar}{z2,s1} 
\fmfdot{z1}
\fmfblob{4thick}{s2,z2}
\fmflabel{{\bf 1}}{s2}
\fmflabel{{\bf 2}}{s1}
\end{fmfgraph*}}
\quad
=
\quad
\rho\,\int 
d{\bf p}\,d{\bf q}\,
\,t^2\left({\bf k}-{\bf p}\right)
\,t\left(q\right)
\,G_0\left(p\right) 
\,G_0\left(q\right) 
\end{equation}

\subsubsection*{DDO:}

\begin{equation}\label{DDO_kr}
\parbox{4.5cm}{
\begin{fmfgraph*}(4,3) \fmfpen{thick}
\fmfsurround{d1,z1,z2,s1,s2}
\fmf{scalar}{s2,s1} 
\fmf{heavy}{s2,z1} 
\fmf{scalar,left}{z1,s2}
\fmf{heavy}{z1,z2} 
\fmf{fermion}{z2,s1} 
\fmfdot{z1,z2}
\fmfblob{4thick}{s1,s2}
\fmflabel{{\bf 1}}{s2}
\fmflabel{{\bf 2}}{s1}
\end{fmfgraph*}}
\quad
=
\quad
\rho\,\int 
d{\bf p}\,d{\bf q}\,
\,t\left({\bf k} - {\bf p}\right)
\,t\left(p\right)
\,t\left({\bf p} - {\bf q}\right)
\,G_0\left(p\right) 
\,G_0\left(q\right) 
\end{equation}

\subsubsection*{DDD:}

\begin{equation}\label{DDD_kr}
\parbox{4.5cm}{
\begin{fmfgraph*}(4,3) \fmfpen{thick}
\fmfsurround{d1,z1,z2,s1,s2}
\fmf{scalar}{s2,s1} 
\fmf{heavy}{s2,z1} 
\fmf{scalar,left}{z1,s2}
\fmf{heavy}{z1,z2} 
\fmf{scalar}{z2,s1} 
\fmfdot{z1}
\fmfblob{4thick}{z2,s2}
\fmflabel{{\bf 1}}{s2}
\fmflabel{{\bf 2}}{s1}
\end{fmfgraph*}}
\quad
=
\quad
-\rho\,\int 
d{\bf p}\,d{\bf q}\,
\,t^2\left({\bf k} - {\bf p}\right)
\,t\left({\bf p} - {\bf q}\right)
\,G_0\left(p\right) 
\,G_0\left(q\right) 
\end{equation}

\subsubsection*{Added together:}

\begin{eqnarray}
\lefteqn{\Sigma_\delta^{(2)}\left(k,z\right) \quad\propto}
\\\nonumber
&&
\rho\,\int
d{\bf p}\,d{\bf q}\,
\left[\,t\left({\bf k} - {\bf p}\right) - t\left(p\right)\,\right]
\left[\,t\left({\bf k} - {\bf p} + {\bf q}\right) - t\left({\bf k} - {\bf p}\right)\,\right]
\,\left[\,t\left({\bf p} - {\bf q}\right) - t\left(q\right)\,\right]
\, G_0\left(p\right) 
\,G_0\left(q\right) 
\end{eqnarray}

The first bracket is proportional to $k^2$. Depending on whether $p$ or $q$ is
small, the third or second bracket yields the required additional $p^2$ or
$q^2$, respectively.

Thus we have here too:

\begin{equation}
\lim_{z\to
    0}\,\lim_{k\to 0}\,{\rm
    Im}\left[\,\Sigma_\delta^{(2)}\left(k,z\right)\,\right]\propto z^{3/2}k^2
\end{equation}

\subsection{$\Sigma_\varepsilon^{(2)}\left(k,z\right)$: Irreducible Diagrams $(1 \cdots 1 \cdots 1)$}

Here only 4 cases are possible.

\subsubsection*{OO:}

\begin{equation}\label{OO_1}
\parbox{4.5cm}{
\begin{fmfgraph*}(4,3) \fmfpen{thick}
\fmfsurround{s2,z2,s1,z1}
\fmf{scalar}{s2,s1} 
\fmf{fermion}{s1,z2} 
\fmf{heavy}{z2,s2}
\fmf{heavy}{s2,z1} 
\fmf{fermion}{z1,s1} 
\fmfdot{z1,z2,s2}
\fmfblob{4thick}{s1}
\fmflabel{{\bf 1}}{s1}
\end{fmfgraph*}}
\quad
=
\quad
-\rho\,\int 
d{\bf p}\,d{\bf q}\,
\,t\left(p\right)
\,t\left({\bf p} - {\bf q}\right)
\,t\left(q\right)
\,G_0\left(p\right) 
\,G_0\left(q\right) 
\end{equation}

\subsubsection*{OD:}

\begin{equation}\label{OD_1}
\parbox{4.5cm}{
\begin{fmfgraph*}(4,3) \fmfpen{thick}
\fmfsurround{s2,z2,s1,z1}
\fmf{scalar}{s2,s1} 
\fmf{fermion}{s1,z2} 
\fmf{heavy}{z2,s2}
\fmf{heavy}{s2,z1} 
\fmf{scalar}{z1,s1} 
\fmfdot{z2,s2}
\fmfblob{4thick}{s1,z1}
\fmflabel{{\bf 1}}{s1}
\end{fmfgraph*}}
\quad
=
\quad
\rho\,\int 
d{\bf p}\,d{\bf q}\,
\,t\left({\bf k} - {\bf p}\right)
\,t\left({\bf p} - {\bf q}\right)
\,t\left(q\right)
\,G_0\left(p\right) 
\,G_0\left(q\right) 
\end{equation}

\subsubsection*{DO:}

\begin{equation}\label{DO_1}
\parbox{4.5cm}{
\begin{fmfgraph*}(4,3) \fmfpen{thick}
\fmfsurround{s2,z2,s1,z1}
\fmf{scalar}{s2,s1} 
\fmf{scalar}{z2,s1} 
\fmf{heavy}{z2,s2}
\fmf{heavy}{s2,z1} 
\fmf{fermion}{z1,s1} 
\fmfdot{z1,s2}
\fmfblob{4thick}{s1,z2}
\fmflabel{{\bf 1}}{s1}
\end{fmfgraph*}}
\quad
=
\quad
\rho\,\int 
d{\bf p}\,d{\bf q}\,
\,t\left({\bf k} - {\bf q}\right)
\,t\left(p\right)
\,t\left({\bf p} - {\bf q}\right)
\,G_0\left(p\right) 
\,G_0\left(q\right) 
\end{equation}

\subsubsection*{DD:}

\begin{equation}\label{DD_1}
\parbox{4.5cm}{
\begin{fmfgraph*}(4,3) \fmfpen{thick}
\fmfsurround{s2,z2,s1,z1}
\fmf{scalar}{s2,s1} 
\fmf{scalar}{z2,s1} 
\fmf{heavy}{z2,s2}
\fmf{heavy}{s2,z1} 
\fmf{scalar}{z1,s1} 
\fmfdot{s2}
\fmfblob{4thick}{z1,z2}
\fmflabel{{\bf 1}}{s1}
\end{fmfgraph*}}
\quad
=
\quad
-\rho\,\int 
d{\bf p}\,d{\bf q}\,
\,t\left({\bf k} - {\bf p}\right)
\,t\left({\bf k} - {\bf q}\right)
\,t\left({\bf p} - {\bf q}\right)
\,G_0\left(p\right) 
\,G_0\left(q\right) 
\end{equation}

\subsubsection*{Added together:}

\begin{eqnarray}
\lefteqn{\Sigma_\varepsilon^{(2)}\left(k,z\right) \quad\propto}
\\\nonumber
&&
\rho\,\int
d{\bf p}\,d{\bf q}\,
\left[\,t\left({\bf k} - {\bf p}\right) - t\left(p\right)\,\right]
\left[\,t\left({\bf k} - {\bf q}\right) - t\left(q\right)\,\right]
\,t\left({\bf p} - {\bf q}\right)
\, G_0\left(p\right) 
\,G_0\left(q\right) 
\end{eqnarray}

For order ${\cal O}\left(k^2\right)$, because of (\ref{tkpk}), the first two
brackets muts be proportional to $\propto {\bf kp}$ and ${\bf kq}$,
respectively. For rotational invariance, the last bracket needs to be expanded
and thus provides the additional required ${\bf pq}$ term, to obtain:

\begin{equation}
\lim_{z\to
    0}\,\lim_{k\to 0}\,{\rm
    Im}\left[\,\Sigma_\varepsilon^{(2)}\left(k,z\right)\,\right]\propto z^{3/2}k^2
\end{equation}

With (\ref{sig_2_split}) it follows immediately
\begin{equation}\label{nonan}
\lim_{z\to
    0}\,\lim_{k\to 0}\,{\rm
    Im}\left[\,\Sigma^{(2)}\left(k,z\right)\,\right]\propto z^{3/2}k^2
\end{equation}
\section{Conclusion}

Working out term by term in the second-order self energy
we have convinced ourselves that to this order the nonalytic
behavior (\ref{nonan}), which both leads to Rayleigh-type sound
attenuation and to the correct long-time tail in the analogous
diffusion problem is recovered. We have done this for
$d=3$, but the generalization to any
$d>1$ is straightforward. Our result is in contradiction to the claims
in the publications \cite{parisi}. In these publications
a self-consistent equation for the self-energy is advocated,
which consists in making the first-order result (\ref{sig_1_ex})
self-consistent, i.e. replacing the 0-th-order Green's function
by the full one. Now, in performing a high-density expansion
of this equation one easily convinces oneself that the
corresponding diagrams are
\begin{itemize}
\item the entire sum 
$\Sigma_\alpha^{(2)}(k,z)$;
\item the diagrams 
$OOO$,
$OOD$,
$ODO$,
$ODD$ of 
$\Sigma_\beta^{(2)}(k,z)$, but not the remaining four diagrams;
\item the diagrams 
$DOO$,
$DOD$,
$DDO$,
$DDD$ of 
$\Sigma_\delta^{(2)}(k,z)$, but not the remaining four diagrams.
\end{itemize}
As the partial sums do not give the correct analytic properties,
this is the reason, why the self-consistent scheme
advocated by \cite{parisi} does not lead to Rayleigh scattering.
We shall publish shortly a self-consistent scheme, which includes
Rayleigh scattering.

\end{fmffile}

\begin{thebibliography}{99}
\bibitem{comment1} 
The real-space transition rate will always carry the
  letter~$r$ in the argument. In all other occations, it will refer to the
  Fourier transform.
\bibitem{comment2} 
For better readability,
  the $z$--dependence of $G_0$ will be suppressed in the following.
\bibitem{comment3} 
In the following we shall
not explicitly mention the $q^2$ from the integration.
\end{thebibliography}
\end{document}